\documentclass[journal]{IEEEtran}
\usepackage{graphicx}
\usepackage{subfigure}

\usepackage{multirow}

\usepackage{amsfonts}

\usepackage{mathrsfs}
\usepackage{epsfig}
\usepackage{color}

\newtheorem{lemma}{Lemma}
\newtheorem{theorem}{Theorem}
\newtheorem{corollary}{Corollary}

\def\bea{\begin{eqnarray}}
\def\eea{\end{eqnarray}}
\newcommand{\re}[1]{(\ref{#1})}
\def\ne{ \nonumber \\ }
\def\nn{ \nonumber }
\def\QED{~\rule[-1pt]{5pt}{5pt}\par\medskip}

\begin{document}
%
\title{Time-Varying and Nonlinearly Scaled Consensus of Multiagent Systems: A Generic Attracting Law Approach}
\author{Mingxuan Sun and Xing Li
\thanks{
This work was supported by National Natural Science Foundation of China under Grant 61573320.}
\thanks{
The authors are with the College of Information Engineering, Zhejiang University of Technology,
Hangzhou, 310023, China.
Email addresses: mxsun@zjut.edu.cn (Mingxuan Sun),
1416440872@qq.com (Xing Li).
}
}

\maketitle
\begin{abstract}
This paper presents the
design and analysis of the finite/fixed-time scaled consensus for multiagent systems.
A study on a generic attracting law,
the certain classes of nonlinear systems that admit attractors with
finite/fixed-time convergence,
is at first given for the consensus purpose.
The estimates for the lower and upper bounds on the settling time functions are provided through the two-phase analysis.
The given estimates are initial state dependent, but the durations are finite, without
regarding the values that the initial states take.
According to the generic attracting law,
distributed protocols are proposed
for multiagent systems
with undirected and detail-balanced directed graphs, respectively,
where the scaled strategies,
including time-varying and nonlinear scales,
are adopted.
It is shown that the finite/fixed-time consensus for the multiagent system undertaken can still be achieved,
even though both time-varying and nonlinear scales are taken among agents.
Numerical simulation of two illustrative examples
are given to verify effectiveness of the proposed finite-duration consensus protocols.
\end{abstract}

\begin{IEEEkeywords}
Scales, finite time convergence, initial conditions, multi-agent systems.
\end{IEEEkeywords}

%
\IEEEpeerreviewmaketitle

\section{Introduction}
\label{intro}
In recent years,
the problem of distributed cooperative control of multi-agent systems has aroused considerable attention \cite{lawton03,jadbabaie03,fax04,olfati04,ren05}.
The broad applications include
rendezvous of mobile autonomous robots
\cite{cortes06,dimarogonas07},
vehicle formations
\cite{lawton03,fax04,lafferriere05},
flocking of mobile sensor networks \cite{tanner07}, and so on.
Through the mutual cooperation between agents,
the large and complex system undertaken
could be coped with,
and tasks can be accomplished by each single agent.
The consensus objective is to apply distributed protocols,
that only require interaction information between the local neighborhoods of agents, due to
the limitations of communication bandwidth and
sensor range,
to make the agents reach an agreement on their states.
The early study on consensus problems was found in \cite{vicsek95},
which gave a simple discrete-time model to simulate the emergence of self-organized particle swarm.
In \cite{jadbabaie03}, a theoretical explanation is provided for the consensus behavior of the Vicsek model, and
the convergence analyses for several inspired models are given. In \cite{ren05},
the results in \cite{jadbabaie03} were extended to the case of directed graphs, where information can be exchanged under dynamically changing interaction topologies.
Graph Laplacians are important which
play a crucial rule in stability and convergence analysis of consensus algorithms \cite{fax04,olfati04}.
It is well known that the algebraic connectivity
can measure the convergence rate,
and the convergence performance can be improved by increasing the algebraic connectivity.
The theoretical framework of the
graph Laplacians based consensus of
multiagent systems is systematically provided in
\cite{olfati07}, and references therein.
Note that special attentions were paid to
the design techniques for the systems on directed graphs
\cite{schurmann89,zhang12,li15}.

The convergence speed of consensus protocols
is usually taken as a type of performance assessment,
an indicator of how effectively and efficiently the consensus is achieved.
Many conventional schemes pursue an asymptotic or exponential solution
which is obtained in an infinite time range.
In fact, one would expect that
the system consensus occurs in finite time, and maintain it afterwards.
Moreover, finite-time convergence is
desirable to satisfy special needs,
such as better disturbance rejection and robustness against uncertainties.
Such a specific requirement is particularly interesting that has caught the attention of researchers.
In \cite{cortes061},
the normalized and signed versions
of the gradient descent
flow of a differentiable function were introduced, and it was shown how the proposed nonsmooth gradient flows achieve consensus in finite-time.
A general framework for designing
finite-time semistable protocols in dynamical networks,
was developed in\cite{hui081},
where semistability is the property
whereby every
system solution converges to a limit point that may depend on the initial condition.

Earlier works were found in \cite{wang10,xiao09}, which
provide an effective way to construct consensus protocols by
continuous state feedbacks
and bridge the gap between
asymptotical consensus and discontinuous finite-time consensus.
In the last ten years,
increasing research efforts have been dedicated to
various finite-time consensus problems, e.g.,
for systems
described by double integrators \cite{wang14},
 for systems under the directed and switching topologies
\cite{li14},
and for systems under the time-varying directed topologies with uncertain leader
\cite{lu17}.
In \cite{liu16}, a switching consensus protocol was designed to solve the finite-time weighted-average-consensus problem
for systems on a fixed directed interaction graph.
It should be noted that for the existing finite-time protocol designs,
the settling time function depends on the initial state of the agent undertaken.
The convergence time cannot be pre-specifiable,
as the initial state is not available.
Moreover, it takes a long time for the convergence,
as the initial state is located far away from the attractor.
The finite-time stability was examined in a seminal paper \cite{polyakov12},
for characterizing that
there exists an upper-bound on the settling time function of the adopted two-term attracting law (AL),
where the term {\it fixed-time} was adopted to describe such convergence property.
The works reported in \cite{parsegov13,zuo14,lu16}
showing the early efforts which were made
to apply such stability theory to the consensus designs.
Furthermore, these problems were solved
for second-order systems \cite{fu16,ni17}, and higher-order systems \cite{zuo18}.
The problem of finite/fixed-time cluster synchronization with pinning control was addressed in \cite{liu18}.
The achievements of consensus are shown to be effect
in coping with input delays \cite{ni17},
handling the
output feedback protocol design for second-order systems
without velocity measurement \cite{tian19},
and
addressing robust performance against bounded uncertain disturbances \cite{fu16,hong17,liu19,wei19}.
It was noticed in \cite{sun18},
that for certain ALs, the duration (or the bound of the duration) of the settling time varies with the initial state, and the duration can be exactly calculated for each given initial state.
It was also shown that the expression for the duration bound can be obtained, and it was proved to be finite, even as the initial state approaches infinity.
In this paper,
AL indicates a desired model
capable of finite-time tracking,
by which the dynamics of the closed-loop system is
governed by the desired model.
The AL approach
is closely related to the pole placement technique for linear systems,
and the consensus objective is to realize the model-following purpose.

Scaled consensus of multiagent systems,
proposed in \cite{roy15},
is an important topic that
deserves much more attention than it has received.
Such coordination tasks are required in practice,
taking for example simultaneous coordination of vehicles both in space and on ground,
due to the huge difference between the scales of vehicles¡¯ position and velocity \cite{meng16}.
Related topics are the
weighted-average consensus with constant weights \cite{liu16}, and
with respect to a monotonic function \cite{jiang11},
and bipartite consensus under cooperative and antagonistic interactions \cite{altafini13,meng17}.
For a complex network composed of two subnets, the problem of scale group consensus was addressed in \cite{yu18}.
In \cite{cortes08},
a class of smooth functions was identified,
for which one can synthesize distributed algorithms that achieve consensus.
In \cite{dong16},
a description of the feasible time-varying formation and an explicit expression
of the time-varying reference function were presented.
Scaled consensus, allowing both nonlinear and time-varying scales,
can be formulated in a unified and general manner for consensus.
To the best of our knowledge, however,
none of the studies have explored yet,
which may be of particular interest as assessed in \cite{liu16,jiang11,altafini13,meng17,cortes08,dong16}.

In this paper,
we investigate the scaled consensus, including both nonlinear and time-varying scales, for multi-agent systems, such that the finite/fixed-time consensus is achieved.
Comparing with the existing works, in particular,
the main contributions of this paper lie in:
i. the scaled consensus with time-varying and nonlinear scales,
achieved by all agents in the network;
ii. the protocol designs for multiagent systems on
undirected and directed graphs, respectively; and
iii. a finite/fixed convergent AL that the duration of the settling time function is finite, whatever the values of initial states take.
The novelty of our proposed protocol design is
its combined use of a generic form of AL,
aiming at the improvement of the convergence performance,
whereas the double power AL is usually adopted
in the related consensus schemes.

The rest of the paper is organized as follows.
In Section \ref{problem}, we give a problem  formulation, with
a description about the scaled consensus to be tackled.
The convergence results of a generic AL are given in subsection \ref{gal}, through the lower and upper bound estimates on the settling time functions.
The main results are presented in Section \ref{sec.main}.
The finite/fixed-time consensus protocols are designed
and analyzed, in subsections \ref{undirectedG} and \ref{directedG},
for systems on
undirected and detail-balanced directed graphs, respectively.
More related issues are addressed in Section \ref{discussions}.
The obtained numerical results are presented in Section \ref{simulation}, and
the conclusion is finally drawn in Section \ref{conclusion}.

\section{Problem Formulation}
\label{problem}
Let us consider a weighted undirected or directed graph $G = \{V, E, A\}$ ( or $G(A)$ for short),
in order to model the interaction topology of the network of $N$ dynamic agents,
for which the consensus problem is tackled in this paper.
$V=\{1, 2, \cdots, N\}$ is the vertex set,
where each vertex represents an agent of the network.
$E \subseteq V \times V$ is the set of connected edges.
$A=[a_{ij}] \in \mathbb{R}^{N \times N}$ is the adjacency weight matrix,
where
$a_{ij} >0$ if and only if  $(i,j) \in E$,
otherwise, $a_{ij} = 0$, $i, j = \{1, 2, \cdots, N\}$.
If all nonzero elements of $A$ are 1, we say $G(A)$ is unweighted.
Here assume that no self-loops in the graph, i.e., $(i,i) \notin E$, or $a_{ii}=0$.
For an undirected graph,
if there is a connection between two nodes $i$ and $j$, then
$a_{ij} = a_{ji}>0$, and $A$ is symmetric.
In contrast, for an directed graph,
if there is a connection from nodes $i$ to $j$, then
$a_{ij}>0$, $A$ is not a symmetric matrix.
$G(A)$ is referred to as a weighted directed graph,
if $ i,j \in E \Leftrightarrow a_{ji} > 0$.
The degree matrix of $G(A)$ is a diagonal matrix $D = {\rm diag} \{d_{1},d_{2},....d_{n}\}$,
where the degree $d_{i}$ of node $i$ is defined as $d_{i}=\sum_{j=1,i\neq j}^{N}a_{ij}$.
The Laplacian matrix of graph $G(A)$ is defined as $L_A = D-A$,
which is symmetric.
An undirected graph $G(A)$ is connected if
there exists a path between $i$ and $j$, $i,j \in \{1, 2, \cdots, N\}$,
between any pair of distinct agent $i$ and agent $j$ in $G(A)$.
For a directed graph,
it is considered to have strong connectivity.
if any two different nodes can reach each other.

We consider a group of $N$ agents, which are described by the following first-order differential equation
\bea \label{agents}
\dot x_i = u_i
 \eea
where $i \in \{1,2, \cdots, N\}$, $x_i$ is the state of agent $i$ and $u_i$ the control input protocol to be designed.
The objective of this paper is to find $u_i, i = 1,2, \cdots, N,$ such that the scaled consensus can be achieved for the $N$ agents described by Eq. \re{agents}, under undirected and directed interaction topology, respectively.

A multiagent system is said to achieve the finite-duration scaled consensus, with both time-varying and nonlinear scales, if, for any $x_{i}(0), i \in \{1, 2, \cdots, N\} $, and
$\forall i,j \in \{1, 2, \cdots, N\}$,
\bea \label{D1}
g_{j}(x_{j},t)=g_{i}(x_{i},t),~~~\forall t \geq T
\eea
where $T$ is the duration of the settling time,
$g_{i}(x_{i},t), i =1,2, \cdots, N,$ represent the scaling functions,
which are assumed to be continuous differentiable, and
$ g_i (x_i,t) \neq 0$ and
$\frac{\partial g_i}{\partial x_i}(x_i,t) \neq 0$.

The scaling can be adopted by a separate manner.
The multiple scales takes $s_{i}(t)g_i(x_{i})$ as the scaling functions, which typically satisfies that

i. $s_{i}(t) \neq 0, i =1,2, \cdots, N,$ indicating the time-varying scales, and

ii. $g_i(x_i) \neq 0, i =1,2, \cdots, N,$
are the nonlinear scales, satisfying that $\frac{dg_i(x_i)}{dx_i} \neq 0$.

The additional scales adopts the form of $s_i(t) + g_i(x_i)$, with the appropriate requirements.
The published scaled consensus designs \cite{roy15,meng16} dealt with constant (multiple) scales, where $s_{i}$ takes a constant value and $g_i(x_i,t) = x_i$.
The nonlinear consensus can be handled by choosing $s_{i}(t) =1$ \cite{cortes08}.
Similar to \cite{dong16}, we apply the additional scales to address the problem of time-varying consensus.
The consensus approach of our paper is a unified one,
allowing
the above described scaling, and has suitability for broad applications.
By the finite/fixed-time consensus
we mean that
the duration of the settling time of each closed-loop system is finite,
whatever value the initial state is.
We shall explain this, in Section \ref{gal}, with the definition of the finite/fixed-time stability.

\section{A Generic Attracting Law}\label{gal}
For the purpose of the scaled consensus,
in this section,
the convergence performance of certain nonlinear systems are considered, which admit attractors with finite settling time.

Let us begin with the introduction of the concepts about finite-time stability.
Consider the scalar nonlinear system $\dot x = f(x), f(0)=0$, and $x(0) = x_0$.
Its zero solution, $x=0$, is said to be globally fixed-time stable, if it is globally finite-time stable and the settling time function, $T(x_0)$, is bounded for arbitrary $x_0$ \cite{polyakov12}.
For the concepts on
asymptotic stability, finite-time stability,
and the definition for settling time function,
we refer to literature \cite{haddad08}.
In this paper,
we suggest an attracting law approach,
which is applicable for both finite-time and fixed-time
stable systems.
Note that the settling time function of a finite-time stable system, $T(x_0)$, is continuous
if and only if it is continuous at $x = 0$.
Hence, as the settling time function is continuous, the settling time is finite for a finite $x_0$.
Moreover, for a fixed-time stable system,
an upper bound of $T(x_0)$ exists,
and
the duration of the interval $[0, T(x_0))$ is finite, whatever the value of $x_0$ takes.
In this paper,
we also wish to give a lower bound of $T(x_0)$, which is helpful for determining the duration, when $x_0$ is not available.

The suggested AL approach
is closely related to the pole placement technique for linear systems.
The pole placement technique is usually applied
for a linear system to realize the model-following purpose,
by which the dynamics of the closed-loop system is
governed by the desired model capable of exponentially asymptotically tracking.
In this paper,
AL indicates a desired model
ensuring finite-time tracking,
and the consensus objective is also to realize the model-following purpose.
We can see that the AL approach presented in this paper relies on an extension of the pole placement technique.

For the consensus purpose, we adopt a generic attracting law (GAL), described by
the following differential equation
\bea \label{fda.fastqp}
\dot x = - \rho x
-\kappa_{1} |x|^{\gamma_{1}} {\rm sgn}(x) - \kappa_{2} |x|^{\gamma_{2}} {\rm sgn}(x),
~x(0) = x_0
 \eea
where $\rho, \kappa_1$ and $\kappa_2$ are positive reals,
$\gamma_{1} = q/p$,
$\gamma_{2} = m/n$,
and $q,  p, m, n $ are odd numbers, satisfying that $q < p$ and $n <m$.
The main idea behind the generic use is to accelerate the
convergence rate in different phases, based on the properties
of the power function.
The GAL involves three terms, all in
the form of $x^\gamma$, satisfying that for  $0 < \gamma' < \gamma'' $,
$|x|^{\gamma'}  < |x|^{\gamma''}$, as $|x| > 1$;
and
$|x|^{\gamma'} > |x|^{\gamma''}$, as $|x| < 1$.
These three
terms in GAL indicates all cases of $\gamma$ ( i.e.,
$0< \gamma < 1$, $\gamma = 1$, and $\gamma > 1$),
and in turn specify a generic action for convergence improvement,
according to the following useful properties:

{\bf P-term (the proportion term, with $\gamma = 1$ )}.
The proportional term with $\gamma = 1$ is introduced, in order to speed up the convergence process.
With this term, the AL is usually referred to as a {\it fast-}AL.
When setting $\kappa_{1} = \kappa_{2} =0$, the AL reduces to $\dot x = - \rho x$, assuring that $x$ converges to zero exponentially, as time increases.

{\bf FT-term (the finite-time term, with $0< \gamma_1 < 1$)}.
Due to the term with $0 < \gamma_1 < 1$, the finite-time convergence of the AL is guaranteed. Obviously, the AL fails to achieve finite-time convergence, as $\kappa_1 =0$.
Note that for $\gamma_1' < \gamma_1''$,
$|x|^{\gamma_1'} > |x|^{\gamma_1''}$, as $|x| < 1$.
As such, the smaller the $\gamma_1$, the faster the convergence rate.

{\bf FD-term ( the finite-duration term, with $\gamma_2 > 1$)}.
The term with
$\gamma_2 > 1$ is needed to ensure that the duration bound on the settling time function is finite.
Without this term, namely, $\kappa_2 =0$,
this AL cannot achieve the bounded-duration performance any more.
Note that for $\gamma_2' < \gamma_2''$,
$|x|^{\gamma_2'} < |x|^{\gamma_2''}$, as $|x| < 1$.
Therefore, the lager the $\gamma_2$, the faster the convergence rate.

For the purpose of comparison, we consider the following AL,
a desired model appeared in the related publications,
\bea
\dot x &=&
-\kappa_{1} |x|^{\gamma_{1}}{\rm sgn}(x)
-\kappa_{2} |x|^{\gamma_{2}} {\rm sgn}(x), ~~x(0) = x_0
  \label{fda.qp}
 \eea
This model has only two terms in the right-hand side of \re{fda.qp},
and lacks the proportion term.
This model is very popular and the style is unique, in the context of finite-time consensus.
We have to make a detailed comparison between this model and the GAL, and  clarify the need and the necessity to introduce the GAL,

The results about the mentioned finite/fixed-time convergent ALs are summarized in the following lemmas,
which will be used for the consensus designs to be presented.

\begin{lemma} \label{lem.fda.fastqp}
The origin of \re{fda.fastqp} is finite-time stable,
associated with the property that the duration of the settling time function satisfies,
for $|x_0| \ge 1$,
\bea \label{fastqp.t1t2.b}
&&\frac{{\rm ln} \Big{(}\frac{\left(1+\frac{\kappa_{2}}{(\rho+\kappa_{1})}\right)}
         {\left(|x_0|^{1-\gamma_2}+\frac{\kappa_{2}}{(\rho+\kappa_{1})}\right)}\Big{)}}{(\rho+\kappa_{1}) (\gamma_{2}-1)}+\frac{{\rm ln} \left ( 1+\frac{(\rho+\kappa_{2})}{\kappa_{1}}
 \right )}{(\rho+\kappa_{2})(1-\gamma_{1})}
 \leq   T(x_0) \ne
&\leq & \frac{1}{\rho (1-\gamma_{1})}{\rm ln} \left ( 1+\frac{\rho}{\kappa_1}  \right ) \ne
&&+\frac{1}{\rho (\gamma_{2}-1)}  {\rm ln} \Big( \left (1+\frac{\kappa_2}{\rho}\right )
/\left (|x_0|^{1-\gamma_2}+\frac{\kappa_2}{\rho} \right ) \Big)
\eea
and for $|x_0|<1$,
\bea
&&\frac{{\rm ln} \left (1 + \frac{(\rho+\kappa_{2})|x_0|^{1-\gamma_1}}{\kappa_{1}}
  \right )}{(\rho+\kappa_{2})(1-\gamma_{1})}
\leq  T(x_0) \ne
& \leq &
  \frac{1}{\rho(1-\gamma_{1})}
{\rm ln} \left ( 1+\frac{\rho}{\kappa_1} (|x_0|)^{1-\gamma_1}\right )
 \label{fastqp.t1t2.a}
\eea
\end{lemma}

\begin{lemma} \label{lem.fda.qp}
The origin of \re{fda.qp} is finite-time stable, where the settling time duration satisfies,
for $|x_0| \ge 1$,
\bea
&&\frac{ {\rm ln}\Big(\frac{\left(1+\frac{\kappa_{2}}{\kappa_{1}}\right)}
        {\left(|x_0|^{1-\gamma_2}+\frac{\kappa_{2}}{\kappa_{1}}\right)}\Big)}
         {\kappa_{1} (\gamma_{2}-1)}
+\frac{{\rm ln}\left ( 1+\frac{\kappa_{2}}{\kappa_{1}}
 \right )}{\kappa_{2}(1-\gamma_{1})}
\leq
T(x_0) \ne
&\leq & \frac{1}{\kappa_{1}} \frac{1}{1-\gamma_{1}}
+\frac{1}{\kappa_2} \frac{\left ( 1 -
|x_0|^{1- \gamma_{2} } \right )}{ \gamma_{2} - 1}
\label{eq.t1t2}
\eea
and for $|x_0| < 1$,
\bea
 \frac{{\rm ln} \Big (1 + \frac{\kappa_{2}|x_0|^{1-\gamma_1}}{\kappa_{1}}
  \Big)}{\kappa_{2}(1-\gamma_{1})}
 \leq  T(x_0)
\leq
\frac{1}{\kappa_1}  \frac{1}{1 - \gamma_{1} } |x_0|^{1 - \gamma_{1} }
 \label{eq.t1t2.1}
\eea
\end{lemma}

According to
\re{fastqp.t1t2.b}-\re{fastqp.t1t2.a},
the finiteness of the duration $[0,T(x_0)]$ with respect to $x_0$ is guaranteed,
which can be expressed as,
for $|x_0| \ge 1$,
\bea \label{eq.T.pterm}
&& \frac{{\rm ln} \left ( 1+\frac{(\rho+\kappa_{2})}{\kappa_{1}}
 \right )}{(\rho+\kappa_{2})(1-\gamma_{1})}  \le  T(x_0) \ne
 &\le & \frac{ 1}{\rho (\gamma_{2}-1)}
{\rm ln} \left (  1+ \frac{\rho}{\kappa_2}\right)
 + \frac{1}{\rho (1-\gamma_{1})}
{\rm ln} \left ( 1+\frac{\rho}{\kappa_1} \right)
\eea
for $|x_0| < 1$,
\bea
 0  \le  T(x_0)
 \le  \frac{1}{\rho (1-\gamma_{1})}
{\rm ln} \left ( 1+\frac{\rho}{\kappa_1} \right)
\eea
It is seen from
\re{eq.t1t2} and \re{eq.t1t2.1}
that $T(x_0)$ is finite
for whatever value $x_0$ takes, and
for $|x_0| \ge 1$,
\bea
 \frac{{\rm ln}\left ( 1+\frac{\kappa_{2}}{\kappa_{1}}
 \right )}{\kappa_{2}(1-\gamma_{1})}
 \leq
   T(x_0)
\leq
\frac{1}{\kappa_{1}} \frac{1}{(1-\gamma_{1})}
+\frac{1}{\kappa_2} \frac{ 1}{ (\gamma_{2} - 1)}
 \label{neq.t1t2}
\eea
for $|x_0| < 1$,
\bea
 0\leq T(x_0)\leq
\frac{1}{\kappa_{1}} \frac{1}{(1-\gamma_{1})}
\eea

Whenever $\rho=0$, the system \re{fda.fastqp} becomes the system \re{fda.qp}.
Therefore, the convergence rate of the system is improved due to the introduction of the term $-\rho x$ in \re{fda.fastqp}.

\begin{corollary}
The finite-time stable system \re{fda.fastqp} has smaller upper and lower bounds of the settling time duration than those of system \re{fda.qp}.
\end{corollary}

This paper provides the GAL, 
raised mainly from the requisition for the convergence performance.
It should be noted that the mentioned-above GAL is not new, but
a well-known technique for fasting upon the rate of convergence
of terminal sliding-mode control of dynamic systems,
guidance of missiles, etc.
However, to the best of our knowledge,
previously none of the studies
have yet been performed
to evaluate the convergence performance
of the protocol designs for consensus of multiagent systems, for which the GAL is adopted.

\section{Finite-Time Scaled Consensus}
\label{sec.main}

On the basis of the GAL,
the protocol designs
for finite-time scaled consensus
and the performance analysis of the multiagent systems undertaken are carried out in this section.

\subsection{Scaled consensus on undirected graphs}
\label{undirectedG}
Let us denote by $L_{A}=[l_{ij}] \in \mathbb{R}^{N \times N}$ the Laplacian of $G(A)$, which is defined as
\bea
l_{ij} = \left\{\begin{array}{ll}
\sum_{k=1,k \neq i}^{n}a_{ik}, & i=j \\
-a_{ij},  & i \neq j
   \end{array}
\right.
\eea
and denote the eigenvalues of $L_A$ by
$\lambda_1, \cdots, \lambda_N$, satisfying that
$\lambda_1 \le \cdots \le \lambda_N$.
It always has a zero eigenvalue, i.e.,
$\lambda_1 = 0$, corresponding to the aligned state $\mathrm{1}_{N} = [1, 1, \cdots, 1]^{T}$.
In addition, as $G(A)$ is connected, $\lambda_2 > 0$
\cite{olfati04,ren05}.
Hence,
for a connected graph, $L_{A}$ is positive semi-definite, i.e., all nonzero eigenvalue of $L_{A}$ is positive.
The Laplacian potential is expressed by
\bea x^{T}L_{A}x = \frac{1}{2}\sum_{i=1}^{N}\sum_{j=1}^{N}a_{ij}(x_{j}-x_{i})^{2} \nn
\eea
According to the algebraic connectivity,
\bea x^{T}L_{A}x \geq \lambda_{2}(L_{A})x^{T}x \nn
\eea
if $\mathrm{1}_{N}^{T}x (= \sum_{i=1}^{N} x_i) =  0$.
Furthermore,
for the connected graph $G(A)$,
a slight modification to Lemma 4 in \cite{meng17} can be given as:

\bea \label{keyeq.lambda2}
x^{T} L_{A}^{2} x \geq \lambda_{2}(L_{A}) x^{T}L_{A} x
\eea
for any $x \in \mathbb{R}^{N}$.
For our analysis purpose,
let us denote
$\chi_{g} = [ g_1, \cdots,  g_N]^T$.
The following result is in turn established.
\begin{lemma}   \label{lem.chi}
For the connected graph $G(A)$,
\bea    \label{keyeq.chi}
\chi_{g}^{T} L_{A}^{2} \chi_{g} \geq \lambda_{2}(L_{A}) \chi_{g}^{T}L_{A}\chi_{g}
\eea
\end{lemma}
{\bf Proof.}
Since the undirected $G(A)$ is connected,
$L_A$ is positive semi-definite and symmetric.
There exists a unique positive semi-definite matrix $M$
such that $L_{A}=MM^{T}=M^{2}$ \cite{horn94}.
We note that $M=M^{T}$ and $L_{A}1_{N}=0$.
Then we obtain $1_{N}^{T}M=0$, which implies
$1_{N}^{T} M \chi_{g} =0$.
It follows from \re{keyeq.lambda2} that
\bea
\chi_{g}^{T}L_{A}^{2} \chi_{g} &=& \chi_{sg}^{T}M^{4} \chi_{g}    \ne
 &=&(M \chi_{g})^{T} L_{A} M\chi_{g} \ne
 &\geq &\lambda_{2}(L_{A})(M \chi_{g})^{T}(M \chi_{g}) \ne
 &=&\lambda_{2}(L_{A})\chi_{g}^{T}L_{A}\chi_{g} \nn
\eea
Hence, inequality \re{keyeq.chi} holds.
\QED

We shall present a continuous protocol to solve the
problem of the scaled consensus for the multi-agent system \re{agents},
which interaction topology is modeled by an undirected graph.
The proposed control protocol for agent $i$ is,
\bea \label{CS.u}
u_{i}&=&  \frac{1}{\left ( \frac{\partial g_i}{\partial x_i} \right )}
 \bigg ( \kappa_{1} \left (
\sum_{j=1}^{N}a_{ij}( g_{j}(x_{j},t)- g_{i}(x_{i},t) )\right )^{\gamma_{1}} \ne
&& +\kappa_{2} \left ( \sum_{j=1}^{N}a_{ij} (g_j(x_{j},t)- g_i(x_{i},t)) \right )^{\gamma_{2}} \ne
&& +\rho \sum_{j=1}^{N}a_{ij}( g_{j}(x_{j},t)- g_{i}(x_{i},t) )
 - \frac{\partial g_{i}}{\partial t} \bigg )
\eea
where $\rho, \kappa_1, \kappa_2 >0$,
$\gamma_{1} = q/p$, $\gamma_{2} = m/n$,
$m$, $n $, $p$, and $q $ are odd numbers, satisfying that $m > n$ and $p > q$,
$\frac{\partial g_i}{\partial x_i}$ and $\frac{\partial g_i}{\partial t}$ represent the partial derivatives of $g_i(x_i,t)$
with respect to $x_i$ and $t$, respectively, and
$V (= \frac{1}{2} \chi_{g}^{T} L_{A}\chi_{g})$
is the Lyapunov function candidate we choose.

\begin{theorem}  \label{thm1}
Consider the multiagent system \re{agents} with a connected communication topology.
Then the protocol \re{CS.u} achieves
the finite/fixed-time consensus, for all initial states.
\end{theorem}
{\bf Proof.} By the definition of $L_A$, let us choose
\bea
V = \frac{1}{4}\sum_{i=1}^{N}\sum_{j=1}^{N} a_{ij}(g_j(x_{j},t)-g_i(x_{i},t))^{2} \nn
\eea
Calculating the derivative of $V$ with respect to time
and applying the protocol \re{CS.u}
give rise to
\bea
&&\dot{V} \ne
&=& \sum_{i=1}^{N} \left ( \frac{\partial V}{\partial x_{i}} \dot{x}_{i}
+ \frac{\partial V}{\partial g_{i}} \frac{\partial g_{i}}{\partial t}
\right )  \ne
&= & \sum_{i=1}^{N} \bigg \{
\frac{\partial V}{\partial x_{i}}
\bigg [\frac{1}{\bigg ( \frac{\partial g_i}{\partial x_i} \bigg )}
\bigg ( \kappa_{1} \bigg( \sum_{j=1}^{N}a_{ij} (g_j(x_{j},t)-g_i(x_{i},t)) \bigg )^{q/p} \ne
  && +\kappa_{2} \bigg ( \sum_{j=1}^{N}a_{ij} g_j(x_{j},t)-g_i(x_{i},t)) \bigg )^{m/n} \ne
     && + \rho \sum_{j=1}^{N}a_{ij}(g_j(x_{j},t)-g_i(x_{i},t))
   - \frac{\partial g_i}{\partial t}       \bigg ) \bigg ]
   + \frac{\partial V}{\partial g_{i}} \frac{\partial g_{i}}{\partial t}   \bigg \}
\nn
\eea
Note that
\bea
\frac{\partial V}{\partial x_{i}} &=& - \frac{\partial g_i}{\partial x_i}
\sum_{j=1}^{N}a_{ij}(g_j(x_{j},t)-g_i(x_{i},t)) \ne
\frac{\partial V}{\partial g_{i}} \frac{\partial g_{i}}{\partial t}&=&
-\frac{\partial g_{i}}{\partial t} \sum_{j=1}^{N} a_{ij}(g_j(x_{j},t)-g_i(x_{i},t)) \nn
\eea
which results in
\bea
\dot{V}
&=& - \sum_{i=1}^{N} \bigg \{\frac{\partial g_i}{\partial x_i}\sum_{j=1}^{N} a_{ij} (g_j(x_{j},t)-g_i(x_{i}),t) \ne
  &&  \bigg [\frac{1}{\left ( \frac{\partial g_i}{\partial x_i} \right )} \bigg ( \kappa_{1} \left ( \sum_{j=1}^{N}a_{ij} (g_j(x_{j},t)-g_i(x_{i},t)) \right )^{q/p} \ne
  && +\kappa_{2} \left ( \sum_{j=1}^{N}a_{ij} g_j(x_{j},t)-g_i(x_{i},t)) \right )^{m/n} \ne
     && + \rho \sum_{j=1}^{N}a_{ij}(g_j(x_{j},t)-g_i(x_{i},t))
      - \frac{\partial g_i}{\partial t}       \bigg ) \bigg ] \ne
&&   + \frac{\partial g_i}{\partial t} \sum_{j=1}^{N}a_{ij}(g_j(x_{j},t) -g_i(x_{i},t))  \bigg \}       \ne
&= &-\kappa_{1}\sum_{i=1}^{N} \left ( \sum_{j=1}^{N}a_{ij}(g_j(x_{j},t)-g_i(x_{i},t)) \right )^{\frac{p+q}{p}}    \ne
&& -\kappa_{2}\sum_{i=1}^{N} \left ( \sum_{j=1}^{N}a_{ij}(g_j(x_{j},t)-g_i(x_{i},t)) \right )^{\frac{m+n}{n}}    \ne
  &&  -\rho\sum_{i=1}^{N}\left ( \sum_{j=1}^{N}a_{ij}(g_j(x_{j},t)-g_i(x_{i},t)) \right )^{2} \nn
\eea
Due to that $\frac{q+p}{2p} \in (0,1)$, $\frac{m+n}{2n}\in(1,\infty)$,
\bea  \label{V11}
&& \dot{V} \ne
&\leq & -\kappa_{1}\left( \sum_{i=1}^{N}\left ( \sum_{j=1}^{N}a_{ij}(g_j(x_{j},t)
                -g_i(x_{i},t))\right )^{2} \right)^{\frac{q+p}{2p}} \ne
& -&\kappa_{2}N^{\frac{n-m}{2n}}\bigg \{  \sum_{i=1}^{N} \bigg( \sum_{j=1}^{N}a_{ij}(g_j(x_{j},t)
                -g_i(x_{i},t))\bigg )^{2} \bigg \}  ^{\frac{m+n}{2n}} \ne
       & -& \rho\sum_{i=1}^{N} \left ( \sum_{j=1}^{N}a_{ij}(g_j(x_{j},t)-g_i(x_{i},t)) \right )^{2}
 \eea
To proceed, the following relationship is needed:
\bea \label{V12}
&&\sum_{i=1}^{N} \left (
\sum_{j=1}^{N}a_{ij}(g_j(x_{j},t)
                -g_i(x_{i},t)) \right )^{2} \ne
&&=(-L_{A} \chi_{g})^{T} (-L_{A} \chi_{g}) \ne
&&=\chi_{g}^{T} L_{A}^{2} \chi_{g}
\eea
Substituting \re{V12} into \re{V11},
we obtain
 \bea
 \dot{V} &\leq &  - \rho \chi_{g}^{T}L_{A}^{2}\chi_{g}
 -\kappa_{1}\bigg ( \chi_{g}^{T}
 L_{A}^{2}\chi_{g} \bigg )^{\frac{q+p}{2p}}   \ne
&& -\kappa_{2}N^{\frac{n-m}{2n}}\bigg ( \chi_{g}^{T}
 L_{A}^{2}\chi_{g} \bigg )^{\frac{m+n}{2n}} \nn
 \eea
It follows by Lemma \ref{lem.chi} that
 \bea
\dot{V}&\leq&
  - 2\rho\lambda_{2}(L_{A})V
  -\kappa_{1}\Big(2\lambda_{2}(L_{A})V\Big)^{\frac{q+p}{2p}}            \ne
   && -\kappa_{2}N^{\frac{n-m}{2n}}\Big(2\lambda_{2}(L_{A})V\Big)^{\frac{m+n}{2n}}   \nn
\eea
Defining $\Lambda=\sqrt{V}$ leads to, as $V \neq 0$,
\bea
\dot{\Lambda}&=&\frac{1}{2\sqrt{V}}\dot{V}      \ne
          &\leq &  -2\rho\lambda_{2}(L_{A})V\frac{1}{2\sqrt{V}}
          -\kappa_{1}\Big(2\lambda_{2}(L_{A})V\Big)^{\frac{q+p}{2p}}\frac{1}{2\sqrt{V}}  \ne
       && - \kappa_{2}N^{\frac{n-m}{2n}}\Big(2\lambda_{2}(L_{A})V\Big)^{\frac{m+n}{2n}} \frac{1}{2\sqrt{V}} \ne
& = & -\rho\lambda_{2}(L_{A})\Lambda
   -\kappa_{1}2^{\frac{q-p}{2p}}\lambda_{2}(L_{A})^{^{\frac{q+p}{2p}}}\Lambda^{\frac{q}{p}} \ne
&& -\kappa_{2}2^{\frac{m-n}{2n}}N^{\frac{n-m}{2n}} \lambda_{2}(L_{A})^{^{\frac{m+n}{2n}}}\Lambda^{\frac{m}{n}}  \nn
\eea
Defing $\rho' = \rho\lambda_{2}(L_{A})$,
$\kappa_1' = \kappa_{1}2^{\frac{q-p}{2p}}\lambda_{2}(L_{A})^{^{\frac{q+p}{2p}}}$,
and $\kappa_{2}' = \kappa_{2}2^{\frac{m-n}{2n}} N^{\frac{n-m}{2n}} \lambda_{2}(L_{A})^{^{\frac{m+n}{2n}}}$, we have
\bea
\dot{\Lambda}
& \leq & -\rho' \Lambda
-\kappa_{1}'\Lambda^{\frac{q}{p}}
- \kappa_{2}' \Lambda^{\frac{m}{n}}
\label{ap.keyeq}
\eea
According to \re{ap.keyeq}, and by invoking Lemma \ref{lem.fda.fastqp}, the conclusion follows.
\QED

When setting that $\rho =0$, \re{CS.u} reduces to the double-power  protocol. Namely,
\bea  \label{TP.u}
&&u_{i}=
 \frac{1}{\left ( \frac{\partial g_i}{\partial x_i} \right )}
\Bigg ( \kappa_{1} \bigg ( \sum_{j=1}^{N}a_{ij}
(g_j(x_{j},t)- g_i(x_{i},t)) \bigg )^{\gamma_{1}} \ne
     && +\kappa_{2} \bigg ( \sum_{j=1}^{N}a_{ij}
(g_j(x_{j},t)- g_i(x_{i},t)) \bigg)^{\gamma_{2}}
-\frac{\partial g_i}{\partial t} \Bigg )
\eea
where $\kappa_1, \kappa_2 >0$,
$\gamma_{1} = q/p$,
$\gamma_{2} = m/n$,
$m$, $n $, $p$, and $q $ are odd numbers, satisfying that $m > n, p > q$,
and $V (= \frac{1}{2}\chi_{g}^{T} L_{A}\chi_{g})$.

With the two-term protocol, the scaled consensus result can be presented in
the following theorem.

\begin{theorem} \label{thm.dp}
System \re{agents} with the connected communication topology
achieves the finite/fixed-time consensus for all initial states, under the double-power protocol \re{TP.u}.
\end{theorem}
{\bf Proof.}
It follows when the protocol \re{TP.u} is applied that
\bea
\dot{V}
&= &-\kappa_{1}\sum_{i=1}^{N} \left ( \sum_{j=1}^{N}a_{ij}(g_j(x_{j},t)-g_i(x_{i},t)) \right )^{\frac{p+q}{p}}    \ne
  && -\kappa_{2}\sum_{i=1}^{N} \left ( \sum_{j=1}^{N}a_{ij}(g_j(x_{j},t)-g_i(x_{i},t)) \right )^{\frac{m+n}{n}} \nn
\eea
Noting that \re{V12} holds,
we obtain
\bea
&& \dot{V} \leq \ne
&& -\kappa_{1}\left ( \sum_{i=1}^{N}\left ( \sum_{j=1}^{N}a_{ij}(g(x_{j},t)
                -g(x_{i},t))\right )^{2} \right )^{\frac{q+p}{2p}} \ne
& -&\kappa_{2}N^{\frac{n-m}{2n}}\bigg \{  \sum_{i=1}^{N} \bigg( \sum_{j=1}^{N}a_{ij}(g_j(x_{j},t)
-g_i(x_{i},t))\bigg )^{2} \bigg \}^{\frac{m+n}{2n}} \ne
 &= & -\kappa_{1}\bigg (  \chi_{g}^{T}
 L_{A}^{2}\chi_{g} \bigg )^{\frac{q+p}{2p}}
-\kappa_{2}N^{\frac{n-m}{2n}}\bigg (  \chi_{g}^{T}
 L_{A}^{2}\chi_{g} \bigg )^{\frac{m+n}{2n}}             \nn
 \eea
due to that $\frac{q+p}{2p} \in (0,1)$, $\frac{m+n}{2n}>1$.
By Lemma \ref{lem.chi}, it follows that
 \bea
\dot{V}\leq
-\kappa_{1}\Big(2\lambda_{2}(L_{A})V\Big)^{\frac{q+p}{2p}}    -\kappa_{2}N^{\frac{n-m}{2n}}\Big(2\lambda_{2}(L_{A})V\Big)^{\frac{m+n}{2n}} \nn
\eea
Defining $\Lambda=\sqrt{ V}$ leads to, as $V \neq 0$,
\bea
\dot{\Lambda}&=&\frac{1}{2\sqrt{V}}\dot{V}      \ne
          &\leq &  -\kappa_{1}\Big(2\lambda_{2}(L_{A})V\Big)^{\frac{q+p}{2p}}\frac{1}{2\sqrt{V}}  \ne
       && - \kappa_{2}N^{\frac{n-m}{2n}}\Big(2\lambda_{2}(L_{A})V\Big)^{\frac{m+n}{2n}} \frac{1}{2\sqrt{V}} \ne
& = &
   -\kappa_{1}2^{\frac{q-p}{2p}}\lambda_{2}(L_{A})^{^{\frac{q+p}{2p}}}
  \Lambda^{\frac{q}{p}} \ne
&& -\kappa_{2}2^{\frac{m-n}{2n}}N^{\frac{n-m}{2n}} \lambda_{2}(L_{A})^{^{\frac{m+n}{2n}}}\Lambda^{\frac{m}{n}}  \nn
\eea

Let us define
$\kappa_1' = \kappa_{1} 2^{\frac{q-p}{2p}}\lambda_{2}(L_{A})^{^{\frac{q+p}{2p}}}$
and \\
$\kappa_{2}' = \kappa_{2}2^{\frac{m-n}{2n}}N^{\frac{n-m}{2n}} \lambda_{2}(L_{A})^{^{\frac{m+n}{2n}}}$.
Then
\bea
\dot{\Lambda} \leq &
-\kappa_{1}'\Lambda^{\frac{q}{p}}
- \kappa_{2}'\Lambda^{\frac{m}{n}}
 \label{keyeq.TP1}
\eea

Then the conclusion follows from \re{keyeq.TP1}, by
invoking Lemma \ref{lem.fda.qp}.
\QED

The derivations presented in the proofs for Theorems \ref{thm1}-\ref{thm.dp}
ensure that the positive definite function $\Lambda$ satisfy \re{fda.fastqp}, with the appropriate selection of protocol parameters.
In turn, it is seen that Lemma \ref{lem.fda.fastqp} plays an important role in finalizing the analysis and facilitating the proof.
Consequently, the settling time finite-duration convergence can be
determined,
through the chosen protocol parameters,
given in
\re{ap.keyeq} in Theorem \ref{thm1}, and
\re {keyeq.TP1} in Theorem \ref{thm.dp}.

\subsection{Scaled consensus on directed graphs}
\label{directedG}
Now we address the problem of the nonlinearly-scaled consensus for multi-agent systems on directed graphs.
To this end,
we introduce the definitions for indegree and outdegree.
The indegree and outdegree of node $i$ of  graph $G$
are defined as: $d_{in}(i)=\sum_{j=1}^{n} a_{ij}$, and
$d_{out}(i)=\sum_{j=1}^{n} a_{ji}$.
Node $i$ in graph $G$ is said to be balanced,
if $d_{in}(i)=d_{out}(i)$.
Obviously, the indegree of each node in the undirected graph is equal to its outdegree.
Hence, every undirected graph is balanced.
However,
for a directed graph, because the edges between the nodes are directed,
$(i, j)\in E$ does not induce $(j, i)\in E$,
and the adjacency matrix $A$ is not necessarily symmetrical.
The directed graph is balanced,
only if the degree of each node is equal to its outdegree.

In addition, the detail balance property is helpful.
If a weighted directed graph $G(A)$ satisfies
the detail-balanced condition in weights,
there are some real numbers $p_{i}>0$, $i = \{1, 2, \cdots, N \}$, such that
$p_{i}a_{ij}=p_{j}a_{ji}$ for $\forall i, j \in \{1, 2, \cdots, N \} $.
Here, $p_{1}, p_{2}, \cdots, p_{N}$ are the detailed balance parameters associated to $G(A)$.
The detail balance parameter $p_1, p_2,\cdots, p_N$ is a positive integer (1 is the only common divisor),
and the positive vector $p=\{p_{1}, p_{2}, \cdots, p_{n}\}$ is not unique.
Particularly, if $p_{1} = p_{2 }=\cdots=p_{N}$,
the calculation can be simplified by letting $p_{i}=1, i = 1, 2, \cdots, N$.
Let us denote by $G(\hat{A})=\{V, E, \hat{A}\}$ the mirror of $G(A)$, with the same node set as $G(A)$,
and $L_{\hat{A}}$ the graph Laplacian for $G(\hat{A})$ with adjacency matrix $\hat{A}$.
The graph Laplacian is defined as
$L_{\hat{A}}=[\hat{l}_{ij}]\in \mathbb{R}^{N \times N}$, whose elements are given by
\bea \label{lij}
\hat{l}_{ij} = \left\{
\begin{array}{ll}
\sum_{k=1,k \neq i}^{n}p_{i}a_{ik}, & i=j \\
- p_{i}a_{ij}, & i \neq j
\end{array}
\right.
\eea
Note that the adjacency matrix $\hat{A}=[\hat{a}_{ij}]$ is symmetric,
and its elements $\hat{a}_{ij}=\hat{a}_{ji}=p_{i}a_{ij}>0$.
Since $G(A)$ is assumed to satisfy the detail-balanced condition in weights, $L_{\hat{A}}$ is positive semi-definite and symmetric,
and
$0=\lambda_1(\hat{A}) < \lambda_2 (\hat{A})\le \cdots \le \lambda_N(\hat{A})$, and
 $x^{T}L_{\hat{A}}x \geq \lambda_{2}(L_{\hat{A}})x^{T}x$ when $1_{n}^{T}x=0,\forall x\in R^{n}$.
Therefore,
we obtain similar property to
that described in Lemma \ref{lem.chi}, i.e.,
$\chi_{g}^{T} L_{\hat{A}}^{2} \chi_{g} \geq \lambda_{2}(L_{\hat{A}}) \chi_{g}^{T}L_{\hat{A}}\chi_{g}
$.

The theoretical result of theorem \ref{thm1} can be directly extended by the discussion on
multi-agent systems on directed graphs.
Consider system \re{agents} with connected communication topology, of which
the weighted directed graph $G(A)$ is strongly connected and detail-balanced, and
the graph Laplacian is defined by \re{lij}.
The designed control protocol \re{CS.u}
is applicable, as $a_{ij}$ is replaced by $\hat{a}_{ij}$.
Then with the revised control protocol for agent $i$,
the nonlinearly-scaled finite/fixed-time consensus can be realized.

As for the result of Theorem \ref{thm.dp},
we modify the double-power protocol \re{TP.u},
by replacing $a_{ij}$ with $\hat{a}_{ij}$.
The convergence result can be established with the similar lines to   those for Theorem \ref{thm.dp}.
As such, the double-power protocol can be applied to solve the
problem of the scaled consensus
for the multi-agent system undertaken, which interaction topology is modeled by a detail-balanced directed graph.

\section{Discussions}
\label{discussions}

The obtained results of multi-agent consensus on undirected/directed graphs, addressing the scaling issue,
are mainly due to the convergence properties of the proposed GAL, presented in Section \ref{gal}.
Consequently, the duration of the settling time function of the multi-agent system undertaken, determined by the designed protocol parameters and the network structure,
is independent of the initial condition.
The convergence rate is improved by our approach,
in comparison with the conventional finite-time system approach that one usually adopted.
In the published literature, there are many protocol designs,
in the context of fixed-time consensus.
The performance improvement of such designs can be made
by directly applying the GAL-based approach.

Conventionally, one does not use scaling, i.e., $s_i(t)=1$ and $g_i(x_i)= x_i$. However, we have to use it, when we face the problem of difference scales of agents, e.g.,
huge difference between the agents' position and velocity in space and on ground.
Researches on simple but useful situations were found in \cite{meng16}, where the scaling functions were taken as $g_i(x_i,t) = x_i/s_{i}$, and $s_{i} \in \{1,2,...,n\}$.
However,
a troublesome situation may occur due to the agents's dynamic structure changes.
In order to verify the time-varying and nonlinear scaling being possible,
we adopt nonlinear scales, where
$g_i(x_i,t)$ is usually nonlinear function of $x_i$ and $t$.
The separated scaling functions, $s_i(t) g_i(x_i)$,
indicate a direct extension to the existing ones and would be a useful alternative.

One interesting issue is the cooperative and antagonistic interactions in multi-agent systems \cite{meng17}.
It is shown that the state of all agents can be agreed in the case of the same modulus but different symbols.
We need to take into account and show that
the proposed protocol \re{CS.u} can be modified to achieve the cooperative and antagonistic behavior in finite duration, which
is given as
\bea  \label{s.u}
&&u_{i} \ne
&=& \frac{1}{\left (\frac{\partial g_{i}}{\partial x_i} \right )}
\Bigg ( \rho \sum_{j=1}^{N}a_{ij}( g_{j}(x_{j},t)-\mathrm{sign}(a_{ij}) g_{i}(x_{i},t) )\ne
&&+\kappa_{1} \left (
\sum_{j=1}^{N}a_{ij}( g_{j}(x_{j},t)- \mathrm{sign}(a_{ij})g_{i}(x_{i},t) )\right )^{\gamma_{1}} \ne
&&+ \kappa_{2} \left ( \sum_{j=1}^{N}a_{ij}(g_j(x_{j},t)- \mathrm{sign}(a_{ij})g_i(x_{i},t)) \right )^{\gamma_{2}} \ne
 && - \frac{\partial g_{i}}{\partial t} \Bigg )
\eea

Difference exists between the steady situation of the scaled consensus and the behavior
of implicit systems and/or algebraic loops. The former is adjustable, through changing the
scale, (namely, by designing the $g_i$), while the latter cannot be changed.
Many of the published schemes
by no means adjust the steady-state of the agents. Our scaled design provides one way for it.

The asymmetry of the Laplacian matrix of the directed graph makes it difficult to
choose a suitable Lyapunov function.
The detailed balance condition, given in \cite{schurmann89} and further specified in \cite{wang10,liu16},
paves the way to solve the difficulty.
Such a matrix $L_{\hat{A}}$ is in fact equivalent to a symmetric Laplace matrix formed
in the case of the undirected communication graph.

With strongly connected and detail-balanced topology, the proposed protocols \cite{wang10,liu16} solve the  finite time average consensus.
By adopting the similar technique, in this paper
we deal with the scaled consensus problem.
More related treatments,
to introduce a Lyapunov function suitable for the analysis on general digraphs, can be found,
for instance,
\cite{zhang12,li15}, which deserve further study
for development and application of new scaled consensus techniques.

\section{Numerical Simulation} \label{simulation}

Two numerical examples are provided in this section
to illustrate the effectiveness of the proposed consensus protocols, with different scale settings,
for which the finite/fixed-time convergence performance is characterized.

\begin{figure}[!t]
\centering
\includegraphics[scale=0.5,trim=0 0 0 0]{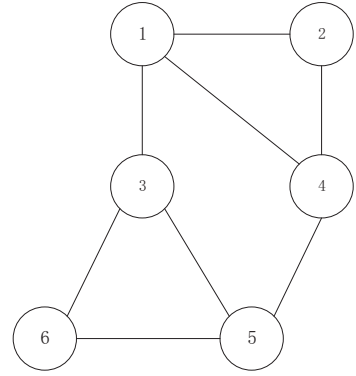}
\caption{Communication topology (Example 1) }
\label{fig1}
\vspace{-10pt}
\end{figure}

\textbf{Example 1:} Consider a group of six agents
whose dynamical behavior is described by \re{agents},
and the interaction topology
is represented by the undirected graph, shown in Fig. \ref{fig1},
with the following adjacency matrix
\bea
A=\left[\begin{array}{cccccc}
0&1&1&1&0&0\\
1&0&0&1&0&0\\
1&0&0&0&1&1\\
1&1&0&0&1&0\\
0&0&1&1&0&1\\
0&0&1&0&1&0
\end{array}\right] \nn
\eea
The initial states of agents are set as
$x_{1}(0)=-18, x_{2}(0)=-8, x_{3}(0)=-5$, $ x_{4}(0)=5, x_{5}(0)=8$, and $x_{6}(0)=18$.

For the numerical simulation,
a separate scaling manner of six agents is taken into account, with
the following two settings, respectively,

Scale setting C1:
\bea
&& g_{1}(x_1,t) = \left( 0.5\mathrm{sin}(2\pi t)+1 \right) \left(5\mathrm{sin}(0.2x_1)+2x_1\right), \ne
&& g_{2}(x_2,t)=\left(-0.5\mathrm{sin}(2\pi t)-1\right)\left(2\mathrm{sin}(0.5x_2)+2x_2\right),\ne
    && g_{3}(x_3,t)=\left(-0.5\mathrm{sin}(2\pi t)-1\right)\left(\mathrm{sin}(x_3)+2x_3\right),\ne
    && g_{4}(x_4,t)=\left(-0.5\mathrm{sin}(2\pi t)-1\right)\left(5\mathrm{cos}(0.2x_4)-2x_4\right),\ne
    && g_{5}(x_5,t)=\left(0.5\mathrm{sin}(2\pi t)+1\right)\left(2\mathrm{cos}(0.5x_5)-2x_5\right), \ne
    && g_{6}(x_6,t)=\left(0.5\mathrm{sin}(2\pi t)+1\right)\left(\mathrm{cos}(x_6)-2x_6\right).\nn
    \eea
Scale setting C2:
\bea
    && g_{1}(x_1,t)=5\mathrm{sin}(0.2x_1)+2x_1, \ne
    && g_{2}(x_2,t)=10\mathrm{sin}(0.5x_2)+10x_2,\ne
    && g_{3}(x_3,t)=\mathrm{sin}(x_3)+2x_3,\ne
    && g_{4}(x_4,t)=-5\mathrm{cos}(0.2x_4)+2x_4,\ne
    && g_{5}(x_5,t)=-10\mathrm{cos}(0.5x_5)+10x_5, \ne
    && g_{6}(x_6,t)=-\mathrm{cos}(x_6)+2x_6. \nn
\eea
Note that the given scales in C1 and C2
satisfy $\frac{dg_{i}(x_{i})}{dx_{i}}\neq 0$.

\begin{figure}
  \centering
  \includegraphics[scale=0.5,trim=0 0 0 0]{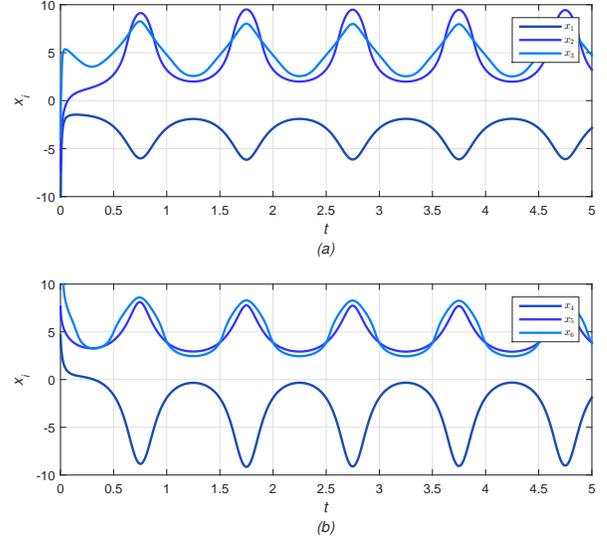}
  \caption{The resultant states under the setting C1:
  (a) states $x_{1}, x_{2}, x_{3}$ and (b) states  $x_{4}, x_{5}, x_{6}$ }
  \label{fig2}
\end{figure}

\begin{figure}
  \centering
  \includegraphics[scale=0.5,trim=0 0 0 0]{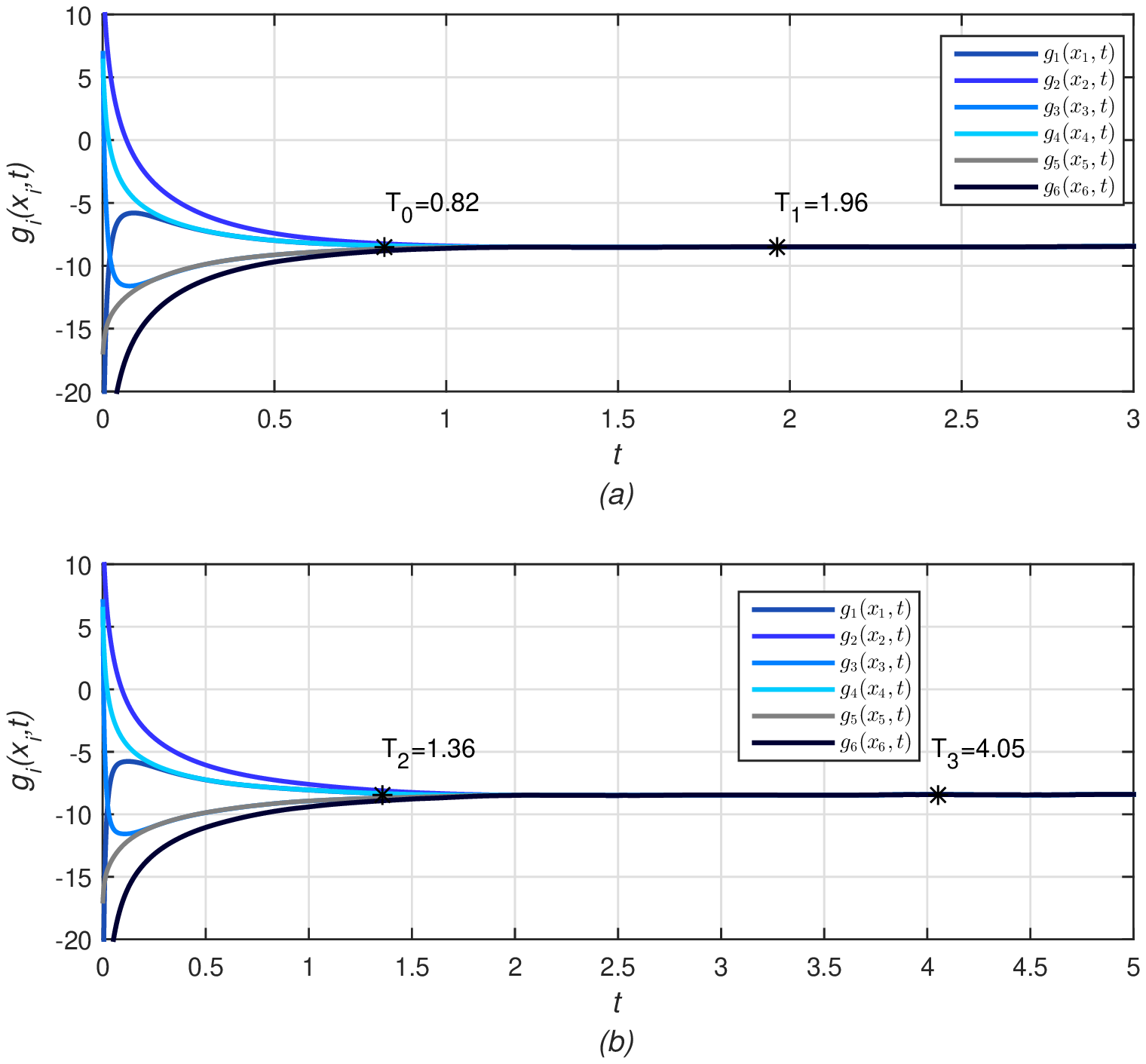}
  \caption{
  The scaled consensus ($g_{i}(x_{i},t)$) under the setting C1:
 (a) by \re{CS.u}  and (b) by \re{TP.u}}
  \label{fig3}
\end{figure}

To illustrate the result of Theorem \ref{thm1},
we apply the protocol \re{CS.u},
with the chosen controller parameters:  $\rho=2, \kappa_{1}=1, \kappa_{2}=1,
\gamma_{1}=\frac{1}{3}, \gamma_{2}=\frac{5}{3}$.
According to the definition of $L_{A}$, the algebraic connectivity of $G(A)$ can be calculated as $\lambda_{2}(L_{A})=1$.
Under the scale setting C1, the numerical results
are shown in Figs. \ref{fig2}-\ref{fig3}.
The resultant states of each multi-agent
are shown in Fig.\ref{fig2}, and
Fig.\ref{fig3} shows the functions $g_{i}(x_{i},t)$, the scaled consensus results by the protocol undertaken.
It is seen from Fig.\ref{fig2}
that the states achieve consensus, according to the different scales, which
show the the scales' impact on the consensus results.
In Fig.\ref{fig3}, we confirm that the protocol design is efficient for
multiagent systems subject to both time-varying and nonlinear scales.
By Lemma \ref{lem.fda.fastqp},
the lower and  upper bounds of the settling time can be calculated as $T_0 = 0.82$, $T_1 = 1.96$,
and from Fig.\ref{fig3} the finite/fixed-time control objective is accomplished.
It exhibits that the settling time is actually
smaller than the upper bound,
and larger than the lower bound of the theoretical estimation.
This is due to the given initial states,
verifying that the actually settling time heavily depends on the the initial states.

\begin{figure}
  \centering
  \includegraphics[scale=0.5,trim=0 0 0 0]{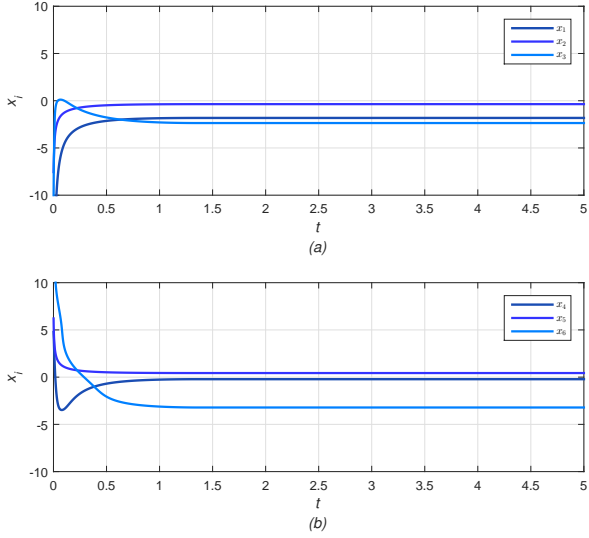}
  \caption{
  The resultant states under the setting C2:
  (a) states $x_{1}, x_{2}$ and (b) states  $x_{3}, x_{4}, x_{5}, x_{6}$
  }
  \label{fig4}
\end{figure}

\begin{figure}
  \centering
  \includegraphics[scale=0.5,trim=0 0 0 0]{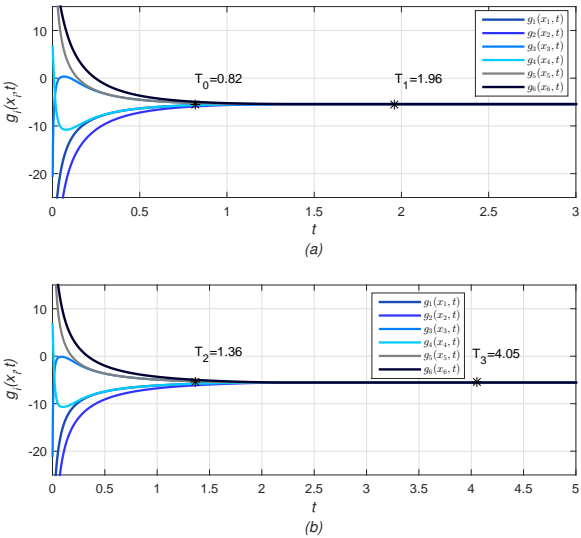}
  \caption{
  The scaled consensus ($g_{i}(x_{i},t)$) under the setting C2:
 (a) by \re{CS.u}  and (b) by \re{TP.u}}
  \label{fig5}
\end{figure}

For comparison,
the simulation is carried out by applying the protocol \re{TP.u},
and the consensus result is shown in Fig.\ref{fig3}.
By Lemma \ref{lem.fda.qp},
the lower and upper bounds of the settling time are calculated as $T_2 = 1.35$, $T_3 = 4.05$,
which gives the reason for the slower convergence rate than that by \re{CS.u},
as shown in Fig.\ref{fig3}.

The simulation results under the scale setting C2
are shown in Figs.\ref{fig4} and \ref{fig5}.
The initial condition of the multiagent undertaken is the same to the above case.
The resultant states of each multi-agent
are shown in Fig.\ref{fig4}, and
the scaled consensus results by the protocol \re{CS.u},
$g_{i}(x_{i},t)$, are given in Fig.\ref{fig5}.
It is observed that the designed consensus protocol achieves the scaled consensus with finite/fixed-time convergence,
leads to the faster convergence rate than that by
applying the protocol \re{TP.u}.

\begin{figure}[!t]
\centering
\includegraphics[scale=0.5,trim=0 0 0 0]{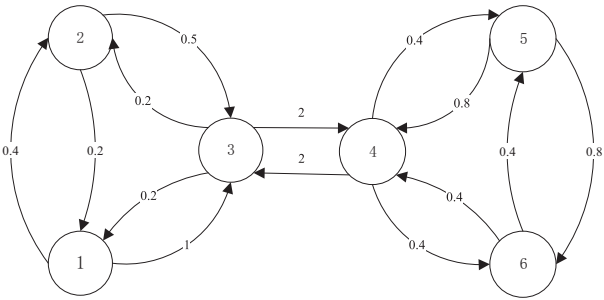}
\caption{Communication topology (Example 2)}
\label{fig6}
\vspace{-10pt}
\end{figure}

\textbf{Example 2:} Consider a six-agent system with the interaction topology modeled as a weighted directed graph,
which adjacency matrix is as follows:
\bea
A=\left[\begin{array}{cccccc}
0   &0.2 &0.2&0&0&0\\
0.4& 0  &0.2&0&0&0\\
1  &0.5 & 0 &2&0&0\\
0  & 0  &2  &0 &0.8 &0.4\\
0  & 0  &0  &0.4 &0 &0.4\\
0 & 0  &0  &0.4&0.8&0
\end{array}\right]\nn
\eea
which is both strongly connected and detail-balanced,
as shown in Fig. \ref{fig6}.
By choosing $p=[10, 5, 2, 2, 4, 2]^{T}$,   $p_{i}a_{ij}=p_{j}a_{ji}$ for $\forall i, j \in \{1, 2, 3, \cdots, N\}$.
We apply the control protocol \re{CS.u},
with $a_{ij}$ being replaced by $\hat{a}_{ij}$, and
the parameter settings:
$\rho=2, \kappa_{1}=1, \kappa_{2}=1,
\gamma_{1}=\frac{1}{3}, \gamma_{2}=\frac{5}{3}$.
The algebraic connectivity of $G(\hat{A})$ can be calculated as $\lambda_{2}(L_{\hat{A}}) = 0.9383$.
In the simulation,
the initial condition are set as $x_0=[-12,-5,-3,12,5,3]$.

For scaling the six agents, the following two sets of time-invariant and time-varying scales are taken into account, respectively.

Scale setting C3:
\bea
&&s_{1}(t)=0.5\mathrm{sin}(2\pi t)+1, s_{4}(t)=-0.5\mathrm{sin}(2\pi t)-1\ne
   && s_{2}(t)=-0.5\mathrm{sin}(2\pi t)-1,   s_{5}(t)=0.5\mathrm{sin}(2\pi t)+1 \ne
   && s_{3}(t)=-0.5\mathrm{sin}(2\pi t)-1,   s_{6}(t)=0.5\mathrm{sin}(2\pi t)+1 \ne
&& g_{i}(x_{i})=x_{i}+\frac{x_{i}}{1+0.1x_{i}^{2}}, i=1,2,3,4,5,6 \nn
\eea

Scale setting C4:
\bea
&& s_{1}(t)= 1, s_{2}(t)= 5, s_{3}(t)=1 \ne
    &&s_{4}(t)=-1, s_{5}(t)=-5, s_{6}(t)=-1  \ne
&& g_{i}(x_{i})=x_{i}+\frac{x_{i}}{1+0.1x_{i}^{2}}, i=1,2,3,4,5,6 \nn
\eea
Note that
$\frac{dg_{i}(x_{i})}{dx_{i}} = 1+\frac{1-0.1x_{i}^{2}}{1+0.1x_{i}^{2}}>0$.

\begin{figure}
  \centering
  \includegraphics[scale=0.5,trim=0 0 0 0]{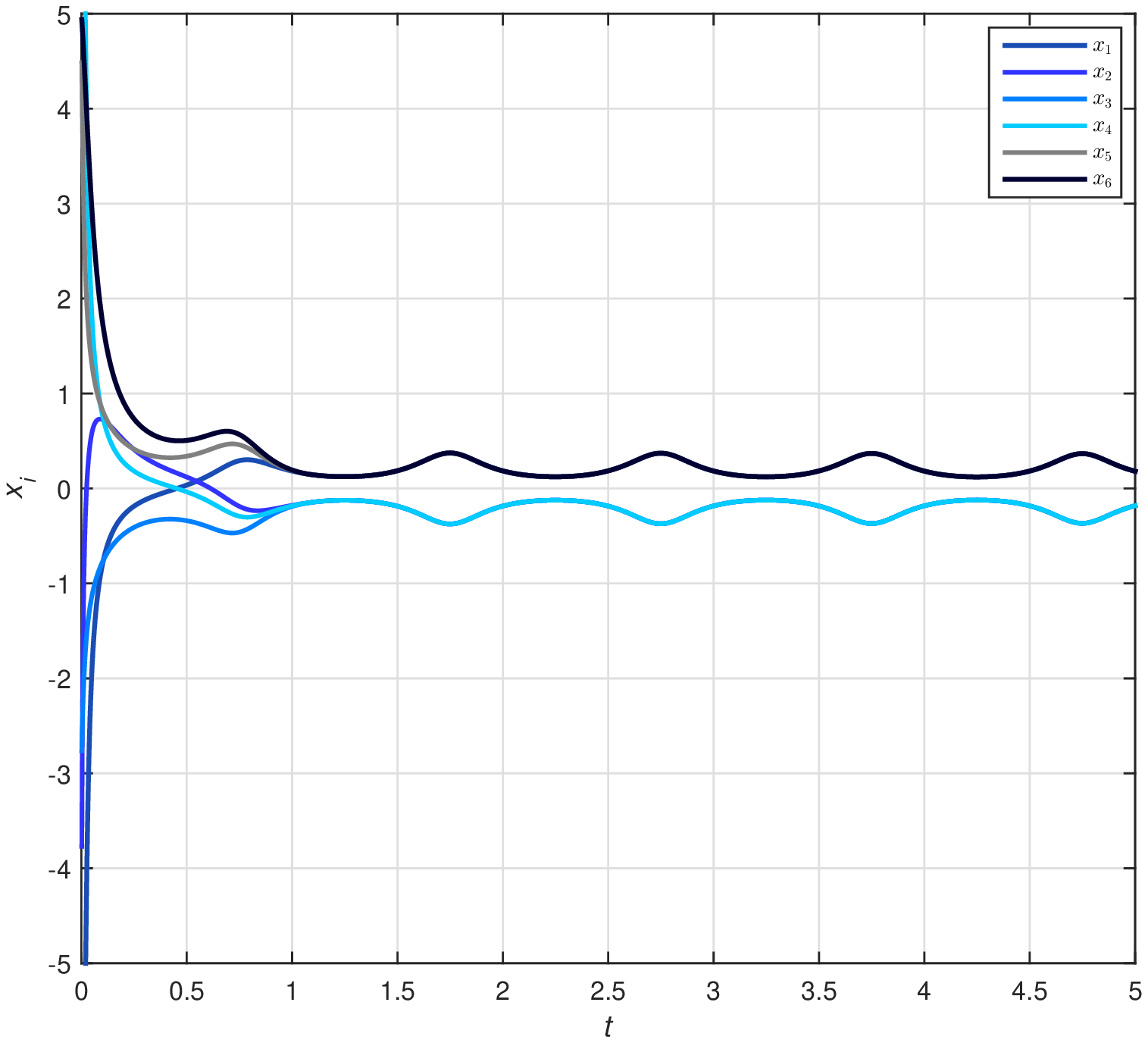}
  \caption{
 The resultant states $x_i, i=1,2,3,4,5,6$ under the setting C3}
  \label{fig7}
\end{figure}

\begin{figure}
  \centering
  \includegraphics[scale=0.5,trim=0 0 0 0]{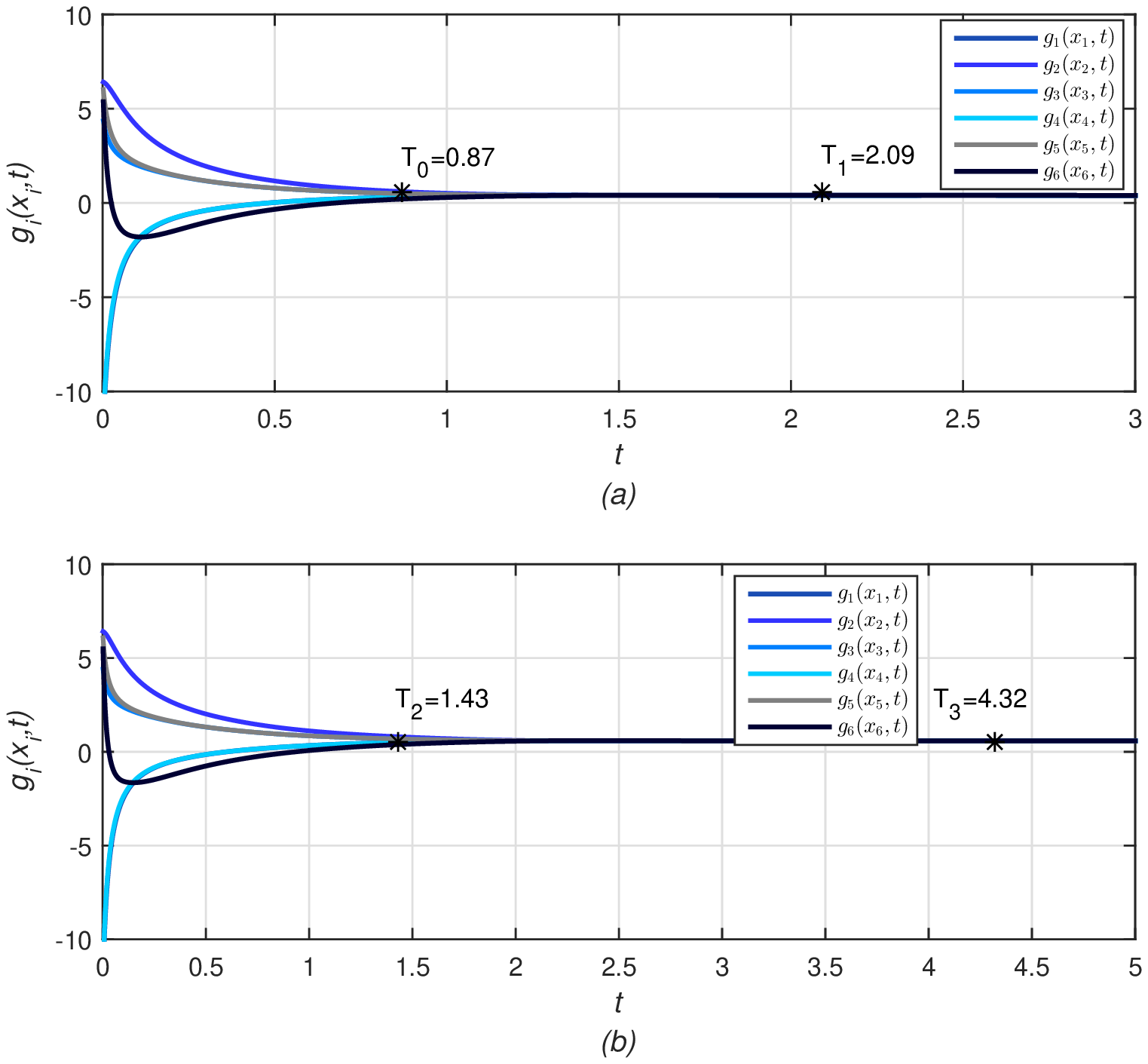}
  \caption{
  The scaled consensus ($g_{i}(x_{i},t)$) under the setting C3:
 (a) by \re{CS.u}  and (b) by \re{TP.u}}
  \label{fig8}
\end{figure}

\begin{figure}
  \centering
  \includegraphics[scale=0.5,trim=0 0 0 0]{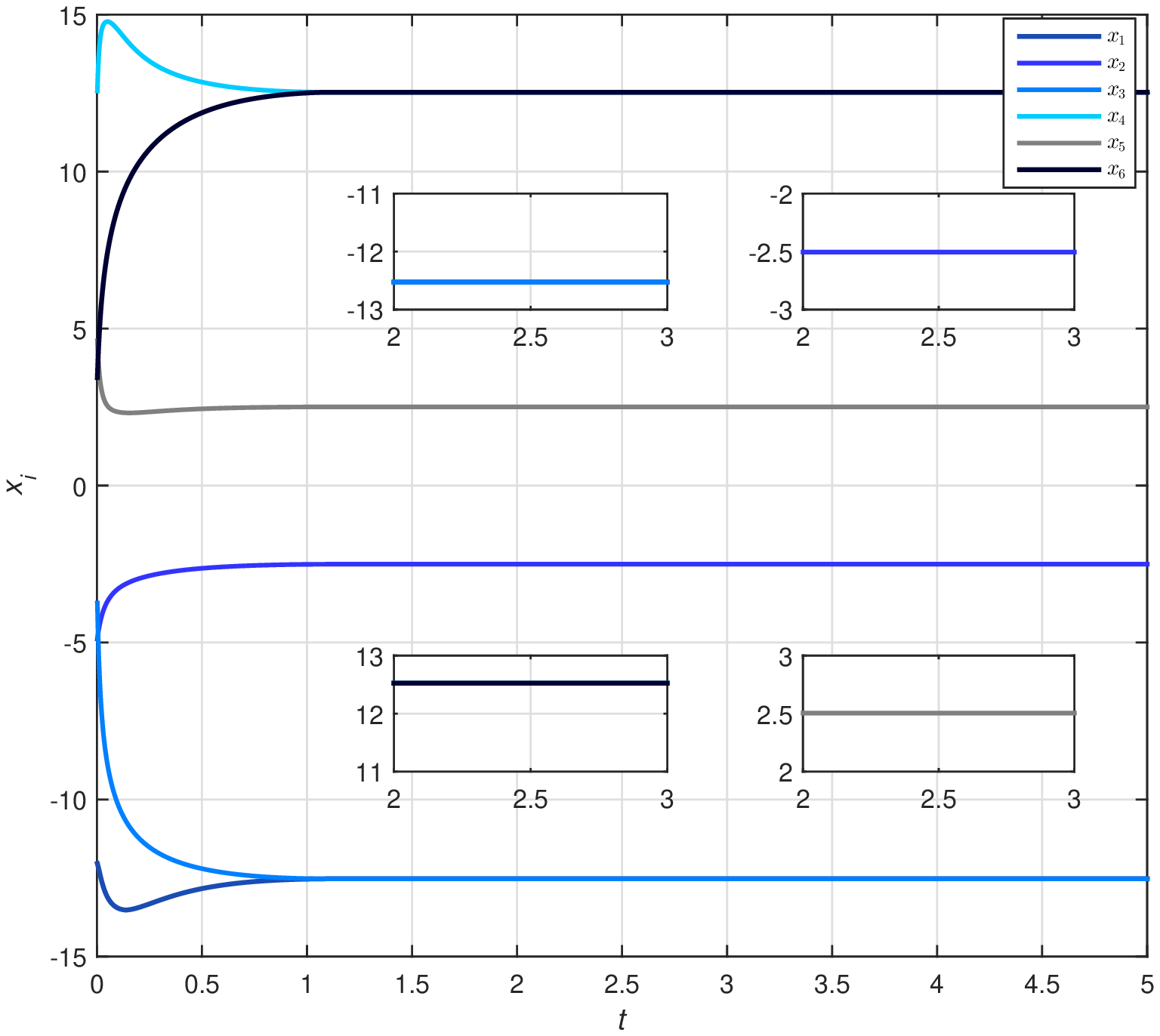}
  \caption{
 The resultant states $x_i, i=1,2,3,4,5,6$ under the setting C3}
  \label{fig9}
\end{figure}

\begin{figure}
  \centering
  \includegraphics[scale=0.5,trim=0 0 0 0]{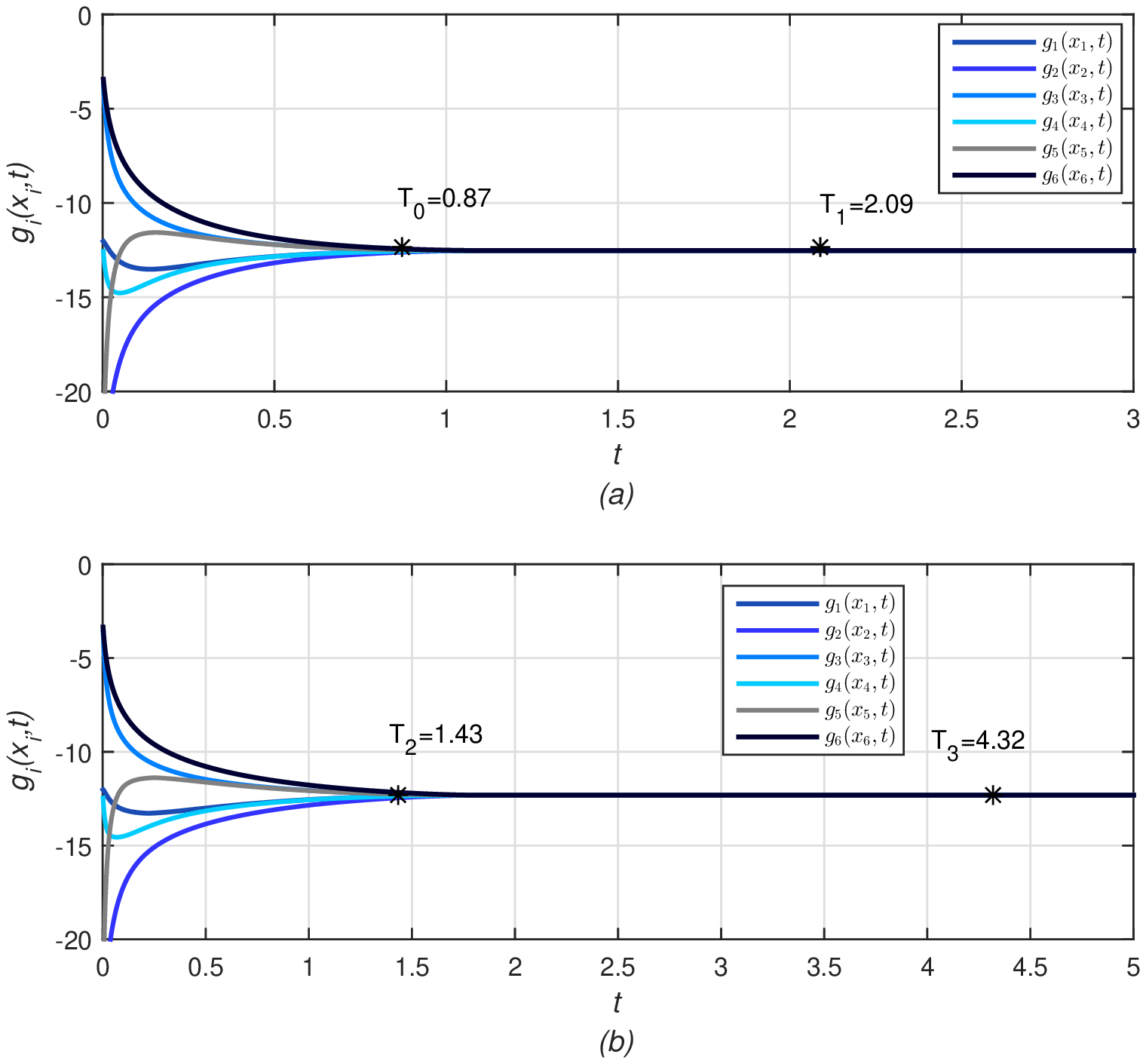}
  \caption{
  The scaled consensus ($g_{i}(x_{i},t)$) under the setting C4:
  (a) by \re{CS.u} and (b) by \re{TP.u}}
  \label{fig10}
\end{figure}

Under scale settings C3 and C4,
we apply the control protocols \re{CS.u} and \re{TP.u},
respectively.
The obtained numerical results are shown in Figs.\ref{fig7}-\ref{fig8} and
Figs.\ref{fig9}-\ref{fig10}, respectively.
Fig.\ref{fig7} and Fig.\ref{fig9} show the resultant state of each agent,
which achieve consensus on different scales.
In Fig.\ref{fig7},
the converged states are time-varying, mainly due to
the time-varying scales,
where two groups of states can be observed.
One is composed of the agents 1, 5 and 6,
the other includes agents 2, 3 and 4.
Two groups of states are opposite in sign.
In Fig.\ref{fig9},
the converged states keep constant, also because of
the time-variant scales,
where
the states of agents 1 and 3 converge to the same value, the states of agents 4 and 6 converge to the other value, and
the converged states of agents 2 and 5 are opposite in sign.
It is seen from Figs. \ref{fig8} and \ref{fig10} that the scaled consensus described by \re{D1} is realized.
Using Lemma \ref{lem.fda.fastqp}, the lower and upper bounds of the settling time can be estimated as $T_0 = 0.87$, $T_1 = 2.09$.
and Using Lemma \ref{lem.fda.qp}, $T_2=1.43$, $T_3=4.32$.

\section{Conclusion}
\label{conclusion}
For the purpose of consensus of multiagent systems, in this paper, stability results of certain class of nonlinear systems have been presented, which admit attractors with finite/fixed-time convergence.
A framework of finite/fixed-time consensus has been provided,
according to the finite/fixed-time stability results,
and distributed protocols are proposed, which realize the scaled consensus, with time-varying and nonlinear scales.
The theoretical analyses of finite/fixed-time convergence have been given
for systems with
undirected and directed graphs, respectively.
The estimates for the upper bounds on the settling time functions are provided, where
the given estimates are initial state dependent, but the durations are finite, without
regarding the values that the initial states take.
It has been shown that the finite/fixed-time scaled consensus for the multiagent system undertaken can be achieved,
despite the adopted time-varying and nonlinear scales.
The simulation results,
showing the desired convergence performance,
have been presented to verify effectiveness of the proposed protocols.


\begin{thebibliography}{99}
\bibitem{lawton03}
J. R. Lawton, R. W. Beard, and B. Young,
``A decentralized approach to formation maneuvers",
{\em IEEE Transactions on Robotics and Automation},
vol. 19, no. 6, pp. 933-941, 2003.

\bibitem{jadbabaie03}
A. Jadbabaie, J. Lin, and A. S. Morse,
``Coordination of groups of mobile autonomous agents using nearest neighbor rules",
{\em IEEE Transactions on Automation Control},
 vol. 49, no. 9, pp. 988-1001, 2003.

\bibitem{fax04}
J. A. Fax and R. M. Murray,
``Information flow and cooperative control of vehicle formations",
{\em IEEE Transactions on Automatic Control}, vol. 49, no. 1, pp. 115-120, 2004.

\bibitem{olfati04}
 R. Olfati-Saber and R. M. Murray,
``Consensus problems in networks of agents with switching topology and time-delays",
{\em IEEE Transactions on Automatic Control}, vol. 49, no. 9, pp. 1520-1533, 2004.

\bibitem{ren05}
W. Ren and R. W. Beard,
``Consensus seeking in multiagent systems
under dynamically changing interaction topologies",
{\em IEEE Transactions on Automatic Control}, vol. 50, no. 5, pp. 655-661, 2005.

\bibitem{cortes06}
 J. Cortes, S. Martinez, and F. Bullo,
``Robust rendezvous for mobile autonomous agents via proximity graphs in arbitrary dimensions",
{\em IEEE Transactions on Automatic Control},
vol. 51, no. 8, pp. 1289-1298, 2006.

\bibitem{dimarogonas07}
D. V. Dimarogonas and K. J. Kyriakopoulos,
``On the rendezvous problem for multiple nonholonomic agents",
{\em IEEE Transactions on Automatic Control},
vol. 52, no. 5, pp. 916-922, 2007.

\bibitem{lafferriere05}
G. Lafferriere, J. S. Caughman, J. J. P. Veerman, and A. Williams,
``Decentralized control of vehicle formations",
{\em Syst and Control Letters},
 vol. 54, no. 9, pp. 899-910, 2005.

\bibitem{tanner07}
H. G. Tanner, A. Jadbabaie, and G. J. Pappas,
``Flocking in Fixed and Switching Networks",
{\em IEEE Transactions on Automatic Control},
vol. 52, no. 5, pp. 863-868, 2007.

\bibitem{vicsek95}
 T. Vicsek, A. Czirok, and E. B. Jacob,
``Novel type of phase transition in a system of self-driven particles",
{\em Physical Review Letters},
vol. 75, no. 6, pp. 1226-1229, 1995.

\bibitem{olfati07}
R. Olfati-Saber, J. A. Fax, and R. M. Murry,
``Consensus and cooperation in networked multi-agent systems",
{\em Proceedings of the IEEE}, vol. 95, no. 1, pp. 215-233, 2007.

\bibitem{schurmann89}
B. Schurmann,
``Stability and adaptation in artificial neural systems",
{\em Physical Review A},
 vol. 40, no. 5, pp. 2681-2688, 1989.

\bibitem{zhang12}
H. Zhang,  F. L. Lewis, and Z. Qu,
``Lyapunov, adaptive, and optimal design techniques for cooperative systems on directed communication graphs", {\em IEEE Transactions on Industrial Electronics},
vol. 59, no. 7, pp. 3026-3041, 2012.

\bibitem{li15}
 Z. Li,  G. Wen, and Z. Duan,
``Designing fully distributed consensus protocols for linear multi-agent systems with directed graphs",
{\em IEEE Transactions on Automatic Control},
vol. 60, no. 4, pp. 1152-1157, 2015.

\bibitem{cortes061}
J. Cortes,
 ``Finite-time convergent gradient flows with applications to network consensus",
{\em Automatica},
 vol. 42, no. 11, pp. 1993-2000, 2006.

\bibitem{hui081}
Q. Hui, W. M. Haddad, and S. P. Bhat,
``Finite-time semistability and consensus for nonlinear dynamical networks",
 {\em IEEE Transactions on Automatic Control},
 vol. 53, no. 8, pp. 1887-1990, 2008.

\bibitem{wang10}
L. Wang and F. Xiao,
``Finite-time consensus problems for networks of dynamic agents",
 {\em IEEE Transactions on Automatic Control},
vol. 55, no. 4, pp. 950-955, 2010.

\bibitem{xiao09}
F. Xiao, L. Wang, J. Chen, and Y. Gao,
``Finite-time formation control for multi-agent systems",
 {\em Automatica},
vol. 45, no. 11, pp. 2605-2611, 2009.


\bibitem{wang14}
X. Wang, S. Li, P. Shi,
``Distributed finite-time containment control for double-integrator multiagent systems",
{\em IEEE Transactions on Cybernetics},
vol. 44, no. 9, pp. 1518-1528, 2014.

\bibitem{li14}
 C. Y. Li and  Z. H. Qu,
 ``Distributed finite-time consensus of nonlinear systems under switching topologies",
{\em Automatica},
vol. 50, no. 6, pp. 1626-1631, 2014.

\bibitem{lu17}
X.~Lu, Y.~Wang, X.~Yu, et al.,
 ``Finite-time control for robust tracking consensus in MASs with an uncertain leader",
{\em IEEE Transactions on Cybernetics},
 vol. 47, no. 5, pp. 1210-1223, 2017.

\bibitem{liu16}
X.~Liu,  J.~Lam,  W.~Yu, et al.,
``Finite-time consensus of multiagent systems with a switching protocol",
{\em IEEE Transactions on Neural Networks and Learning Systems}, vol. 27, no. 4, pp. 853-862, 2016.

\bibitem{polyakov12}
A. Polyakov,
Nonlinear feedback design for fixed-time stabilization of linear control systems,
{\em IEEE Transactions on Automatic Control},
vol. 57, no. 8, pp. 2106-2110, 2012.

\bibitem{parsegov13}
S. Parsegov, A. Polyakov, and P. Shcherbakov,
``Fixed-time consensus algorithm for multi-agent systems with integrator dynamics",
{\em IFAC Proceedings Volumes},  vol. 46, no.27, pp. 110-115, 2013.

\bibitem{zuo14}
Z. Zuo and L. Tie,
``A new class of finite-time nonlinear consensus
protocols for multi-agent systems",
{\em  International Journal of Control}, vol. 87, no. 2, pp. 363-370, 2014.

\bibitem{lu16}
W. Lu, X. Liu, and T. Chen,
``A note on finite-time and fixed-time stability",
{\em Neural Networks}.
vol. 81, pp. 11-15, 2016.

\bibitem{fu16}
J. Fu and J. Wang,
``Fixed-time coordinated tracking for second-order multi-agent systems with bounded input uncertainties",
{\em Systems. Control Letters}.
 vol. 93, pp. 1-12, 2016.

\bibitem{ni17}
 J. Ni,  L. Ling,  C. Liu, et al.,
``Fixed-time leader-following consensus for second-order multiagent systems with input delay",
{\em IEEE Transactions on Industrial Electronics}.
vol. 64, No. 11, pp. 8635-8646, 2017.

\bibitem{zuo18}
 Z. Zuo,  B. Tian,  M. Defoort, et al.,
``Fixed-time consensus tracking for multiagent systems with high-order integrator dynamics",
{\em IEEE Transactions on Automatic Control}.
vol. 63, No. 2, pp. 563-570, 2018.

\bibitem{liu18}
X. Liu, T. Chen,
``Finite-Time and Fixed-Time Cluster Synchronization With or Without Pinning Control",
IEEE Transactions on Cybernetics,
vol. 48, no. 1, pp. 240-252. 2018.

\bibitem{tian19}
B.~Tian,  H.~Lu, Z.~Zuo, et al.,
``Fixed-time leader-follower output feedback consensus
for second-order multiagent systems",
{\em IEEE Transactions on Cybernetics},
vol. 49, no. 4, pp. 1545-1550, 2019.

\bibitem{hong17}
 H.~Hong,  W.~Yu,  G.~Wen, et al.,
``Distributed robust fixed-time consensus for nonlinear and disturbed multiagent systems",
{\em IEEE Transactions on Systems, Man, and Cybernetics: Systems},
vol. 47, no. 7, pp. 1464-1473, 2017.

\bibitem{liu19}
X. Liu, D. W. C. Ho, Q. Song, et al.,
``Finite/Fixed-Time pinning synchronization of complex networks with stochastic disturbances",
{\em IEEE Transactions on Cybernetics}, vol. 49, no. 6, pp. 2398-2403, 2019.

\bibitem{wei19}
 X.~Wei,  W.~Yu,  H.~Wang, et al.,
``An observer based fixed-time consensus control
for second-order multi-agent systems with disturbances",
{\em IEEE Transactions on Circuits and Systems II: Express Briefs}.
vol. 66, No. 2, pp. 247-251, 2019.

\bibitem{sun18}
 M.~Sun,
``Two-phase attractors for finite-duration consensus of multiagent systems",
{\em IEEE Transactions on Systems, Man, and Cybernetics: Systems}, 2017, doi:10.1109/TSMC.2017.2785314

%
\bibitem{roy15}
S.~Roy,
``Scaled consensus ",
{\em Automatica}, vol. 51, pp. 259-262, 2015.

\bibitem{meng16}
D.~Meng and Y.~Jia,
``Robust consensus algorithms for multiscale coordination control of multivehicle systems with disturbances",
{\em IEEE Transactions on Industrial Electronics},
vol. 63, no. 2, pp. 1107-1119, 2016.


\bibitem{jiang11}
 F.~Jiang and L.~Wang,
``Finite-time weighted average consensus with respect to a monotonic function and its application",
{\em Systems \& Control Letters},
vol. 60, pp. 718-726, 2011.

\bibitem{altafini13}
C.~Altafini,
``Consensus problems on networks with antagonistic interactions",
{\em IEEE Transactions on Automatic Control},
vol. 58, no. 4, pp. 935-946, 2013.

\bibitem{meng17}
D.~Meng,  Y.~Jia and J.~Du,
``Finite-time consensus for multiagent systems with cooperative and antagonistic interactions",
{\em IEEE Transactions on Neural Networks and Learning Systems}, vol. 27, no. 4, pp. 762-770, 2017.

\bibitem{yu18}
J. Yu, Y. Shi,
``Scaled group consensus in multiagent systems with first/second-order continuous dynamics",
{\em IEEE Transactions on Cybernetics},
vol. 48, no. 8, pp. 2259-2271, 2018.

\bibitem{cortes08}
 J.~Cortes,
``Distributed algorithms for reaching consensus on general functions",
{\em Automatica},  vol. 44, pp, 726-737, 2008.

\bibitem{dong16}
 X.~Dong and G.~Hu,
``Time-varying formation control for general linear multi-agent
systems with switching directed topologies",
{\em Automatica},  vol. 73, pp. 47-55, 2016.


\bibitem{haddad08}
 W. M. Haddad and  V. Chellabonia,
``Nonlinear Dynamical Systems and Control: A Lyapunov-Based Approach",
{\em New Jersey: Princedon University Press}, 2008.

\bibitem{horn94}
 R. A. Horn and  C. R. Johnson,
``Matrix Analysis",
{\em Cambridge, U. K.: Cambridge University Press}, 1985.

\end{thebibliography}
\end{document}